# SECURING VIRTUALIZED DATACENTERS

Timur Mirzoev, Georgia Southern University; Baijian Yang, Ball State University

## Abstract

Virtualization is a very popular solution to many problems in datacenter management. It offers increased utilization of existing system resources through effective consolidation, negating the need for more servers and additional rack space. Furthermore, it offers essential capabilities in terms of disaster recovery and potential savings on energy and maintenance costs. However, these benefits may be tempered by the increased complexities of securing virtual infrastructure. Do the benefits of virtualization outweigh the risks? In this study, the authors evaluated the functionalities of the basic components of virtual datacenters, identified the major risks to the data infrastructure, and present here several solutions for overcoming potential threats to virtual infrastructure.

## Introduction

The past few years have seen a steady increase in the use of virtualization in corporate datacenters. Virtualization can be described as the creation of computer environments within operating systems allowing multiple virtual servers to run on a single physical computer. Many datacenters have adopted this architecture to increase the utilization of available system resources, which resulted in a decreased need for additional servers and rack space. There are substantial savings in energy, cooling, administration and maintenance expenses. But these advantages are somewhat tempered by the additional complexities, performance and security complications in virtualization environments. According to a recent survey from 531 IT professionals [1], concerns about security was considered one of the top issues in adopting virtual technology (43%) and the main reason that organizations were slow in the deployment of virtualization (55%). It is therefore necessary to examine the security risks associated with virtualization and the potential countermeasures.

Presented here are the key components that make up a virtual datacenter, followed by the analysis of virtualization security threats including virtual machine security, hypervisor security, network storage concerns, and virtual center security. Finally, the authors present possible solutions, guidelines and recommendations for enhancing security in virtualized datacenters.

## Components of Virtual Datacenters

Virtual technologies have a broad range of contexts: operating Systems (OS), programming languages and computer architecture [12]. Virtualization of OS and computer architecture significantly benefits any disaster-recovery process and improves business continuity due to the fact that images of virtual machines can be quickly restored from different physical servers without waiting on the hardware repair. A virtual datacenter is a type of infrastructure that allows for sharing physical resources of multiple physical servers across an enterprise. This aggregation of resources becomes feasible when a suite of virtualization software is installed on the various components such as physical servers, network storage and others. As shown in Figure 1, the major elements of the virtualized datacenter include virtual machines, hypervisors, network resources, and datastores – Network Attached Storage (NAS), Storage Attached Network (SAN) and IP Storage Attached Network (IP SAN).

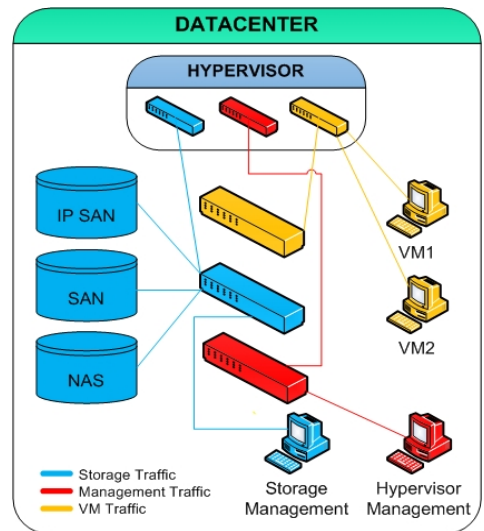

**Figure 1. Datacenter components**

It should be noted that this discussion of virtual infrastructure is independent of specific brands of virtualization suites. Different virtualization venders tend to use different names when referencing the various components, although they all equate to the same basic principles.

## A. Virtual Machines

The concept of virtualization is not new. It is based on a time-sharing concept originally developed by scientists at the Massachusetts Institute of Technology (MIT) in 1961 [2]. Time-sharing creates an opportunity for concurrently managing multi-hardware and multi-user environments on a single physical machine. Today, many vendors such as IBM, VMware, Oracle, HP and others have taken this time-sharing concept even further and developed virtualization schemes of various types including Integrity VM by HP [5]. The advantages of modern technologies such as Integrity VM allow any operating system to run inside VM that supports the Integrity VM platform. An example of how virtualization is applied in a hosted virtualization model is shown by Figure 2.

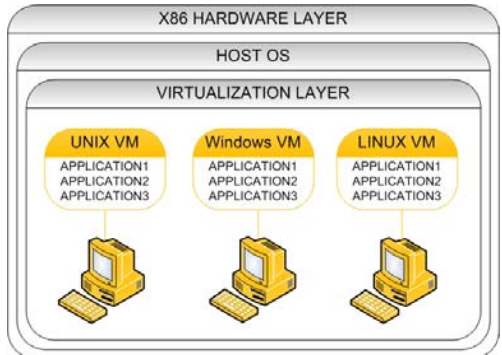

**Figure 2. Hosted approach for virtual servers**
Source: Adapted from [4]

Virtual machines (VMs) are sets of files that support operating systems and applications independent of a host operating system. Typical VMs include BIOS, configuration and disk files. In other words, they are software-only implementations of computers that execute programs just like conventional systems. Figure 3 lists a number of examples of VMs containing virtual hardware components such as CPU, memory, hard drives, network adapters and many others.

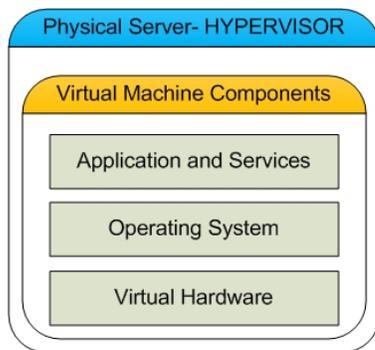

**Figure 3. Elements of Virtual Machines**

Nowadays, the detection of a virtualized OS is a fairly new technique, so most virtual OS environments are not detected by common operating systems such as Windows XP, Vista, and a few Linux operating systems. This offers a great deal of flexibility in the deployment of virtual infra-structures. Another essential property of virtual machines is OS isolation – if a virtual computer gets infected with a virus other virtual machines should not get infected. Popek and Goldberg [8] describe a virtual machine as "an efficient, isolated duplicate of a real machine."

It is possible to run multiple virtual machines on a single physical server without any interaction between operating systems. Thus, if a virtual machine crashes, it would not affect the performance of the other virtual machines on the same physical server. Because of these properties, virtual machines are considered one of the building blocks of virtualized datacenters.

## B. Hypervisors

A hypervisor, also known as physical host, is a physical server that contains hardware and virtualization layer software for hosting virtual machines. Hypervisors use existing physical resources and present them to virtual machines through the virtualization layer. Virtualization is neither simulation nor emulation. To share resources, hypervisor runs in a native mode, directly representing the physical hardware. Additionally, hypervisors effectively monitor and administer shared resources given to any virtual machine. Groups of similar hypervisors may be organized into clusters. Clusters allow for aggregation of computing resources in the virtual environment and allow for more sophisticated architectures. Typically, specific software manages a cluster of hypervisors.

## C. Network Configurations

Virtual datacenters, like traditional datacenters, require network infrastructure. A hypervisor can use multiple network interface cards (NICs) to provide connectivity. In fact, it is highly recommended that several NICs be used for a single hypervisor in order to separate networks and various functions. Furthermore, as virtualization of I/O devices takes off, the IT industry will see another sharp turn towards fewer cables, NICs and other devices. Each Ethernet adapter can be configured with its own IP address and MAC address manually or automatically. A hypervisor or a virtual machine can be configured with one or more virtual NICs to access different physical or virtual networks (see Figure 4).

Virtual switches allow virtual machines to communicate with each other using normal TCP/IP protocols without using additional physical network resources internal to hypervisor. Virtual switches may be configured to have the same functionality as a physical switch, with the exception that direct interconnection between virtual switches is not available in some configurations.

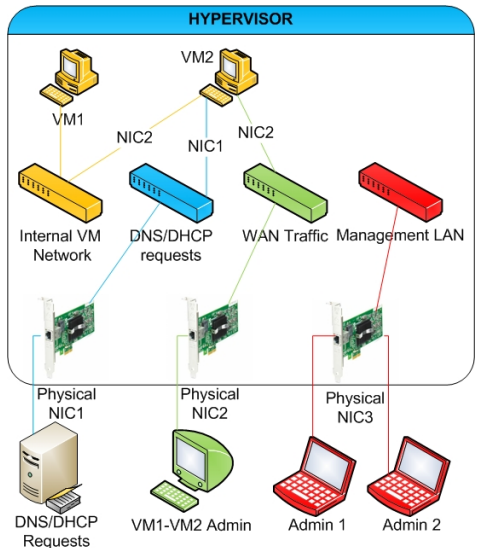

**Figure 4: Hypervisor and virtual machine networking**

## D. Network Storage

Most datacenters rely heavily on Fiber Channel SAN, iSCSI arrays, and NAS arrays for network storage needs. Servers use these technologies to access shared storage networks. Storage arrays are shared among computer resources enabling essential virtualization technologies such as live migration of virtual computers and increased levels of utilization and flexibility. Within a virtual datacenter, network storage devices may be virtualized in order to provide distributed shared access among virtual machines. Moreover, it is recommended that files containing the virtual machines be placed on shared storage. This is because most current virtualization server products will not support a live migration of a virtual machine from one physical server to another physical server if the virtual machine is located on a storage area that is not accessible by all hypervisors.

Several options are available for the creation of datastores such as NAS, SAN, local storage, etc. Shared network storage is an essential component for any virtual infrastructure, and applying different storage technologies typically requires a balance between cost and performance. But regardless of whether it is a NAS or SAN array, securing network storage should always be at the top of the security policy for any datacenter. It is critical to separate network storage traffic from virtual machine traffic.

# Virtualization Security Threats

In this section, the security threats to a datacenter's key components in a virtualized environment are discussed.

## A. Virtual Machine Threats

It is a common misconception that the security risks of virtual machines are much higher than those of physical computers. Virtual machines have the same, if not fewer, levels of security risks as their physical-computer counterparts. This is due to the fact that physical connectivity, software updates and networking employ the same logical infrastructure. The key difference lies in the fact that VMs are running on top of a virtualization layer instead of the actual hardware. Virtual machines should be protected by antivirus software and should be patched on a regular basis just like any physical computer. However, there are several virtualization-specific threats that require the attention of datacenter administrators:

1) **Suspended Virtual Machines**. When a virtual machine is not running, users typically assume there are no security threats since the VM is powered off. Throughout tests in our lab environment, an observation was made that un-patched and unprotected Windows 2000 Servers, running in an earlier version of a virtualized product, were actually infected by the Blaster Worm even when a VM was not running. To make matters worse, it was able to duplicate itself and infiltrate other unprotected suspended virtual Windows 2000 Servers. It was felt that the problem was due to security flaws in how the virtualized networking environment was implemented, but there were no validated hypotheses generated. Another observation was made regarding a virtual machine running as a DHCP server that continued to hand out IP addresses even when it was powered off. It is important then to realize that suspended virtual machines should not be ignored as security targets in virtual datacenter environments.

2) **Resource Contention**. When multiple virtual machines run on the same physical server, they will inevitably be competing for hardware resources such as CPUs, RAMs, I/O, and network. Additionally, all of the server and client operating system updates are typically applied at the exact same time. When multiple virtual machines demand the same physical resources simultaneously, they may cause performance degradation across the datacenter and could even lead to some virtual machines not being able to power on. VM-specific countermeasures must be utilized to alleviate resource contention among virtual machines.

3) **VM Sprawl**. With the help of current virtualization products, the creation of new VMs is quick and easy. It is a common malpractice that whenever there is a need for a certain service, a dedicated VM is immediately created. It then becomes a problem noted as VM Sprawl, where too many virtual machines are created without careful planning, management and service consolidation [9]. The biggest threat associated with VM sprawl could be the cost of resources: storage space, CPU and RAM resources may become scarce and additional software licenses and network service may be required. From a security of point of view, if a virtual datacenter is abused by the creation and deployment of unnecessary virtual machines, the entire infrastructure may become too complex to manage, thereby posing a dangerous, unsecure environment. And if any virtual machine is not carefully protected or it is placed on a wrong network, it may put the entire virtual datacenter in jeopardy.

## B. Hypervisor Threats

Hypervisor is a thin layer that runs directly on top of the physical hardware and provides isolation between different virtual machines that run on the same physical server. It is, therefore, vital to protect hypervisor from being compromised or attacked. Anything happening at the hypervisor level is not visible to virtual machines and will eventually render the traditional OS hardening or protection techniques completely useless. Lately, a number of interesting research efforts have been made to address the possible security threats to hypervisors:

1) **Virtual-Machine-Based Rootkit (VMBR)**. Samuel T. King et al. [11] developed proof-of-concept VMBRs that can be inserted underneath the targeting virtual machines if an attacker gains administrative privileges on a virtual machine. Once the rootkit is successfully installed and configured, it functions as a modified hypervisor on the infected physical server and loads all virtual machines. As a result, an attacker will have a great possibility of controlling every virtual machine that runs on that hypervisor.

2) **Blue Pill Attack**. Joanna Rutkowsaka [6] presented a highly sophisticated attack at the 2006, annual Black Hat Conference. The basic idea of a Blue Pill attack is to exploit the AMD64 SVM virtualization instruction sets—code name Pacifica. If the attack is successful at the chip level, it will also install its own hypervisor to intercept the communication between virtual servers and the hardware. The author also claimed that it is quite likely that similar attacks can be implemented against Intel's VT-x instruction sets—code name Vanderpool. Unlike VMBRs, the Blue Pill attack can be installed on the fly without modifying the BIOS, and there are no noticeable performance penalties associated with the attack. As a result, it becomes extremely difficult to detect such types of attacks.

## C. Virtual Infrastructure Threats

Virtual infrastructures with clusters of hypervisors are highly sensitive to internal attacks. Frequently, the response to internal threats is such that "nothing can be done". That is exactly the reason why internal attacks still exist. If there are no preventive measures that are taken towards internal threats, then internal attacks should be expected. Specifically, the following security threats should be addressed:

1) **Single Point of Control**. A single administrator may be implementing permissions, authentications and privileges to cluster-wide environments of hypervisors, or virtual centers. Such a person becomes the biggest threat to the company's assets, should this super administrator become dissatisfied with the company for any reason.

2) **Physical Access**. If a person gains physical access to a hypervisor, the damage could be much worse because the entire virtual infrastructure can quickly be copied, modified or even removed.

3) **Licensing Server**. Typical virtualization server products are activated and unlocked by the presence of valid license files. If for any reason the license server fails, IT administrators should respond as quickly as possible to get the license server back online. Otherwise, when the grace period expires—for some vendors this can be 14 days—certain features will be disabled, leading to a chaotic environment in the datacenter.

## D. Virtual Network Threats

From a design point of view, there are no essential differences between virtual and physical networking. If network administrators have proven networking skills in dealing with physical networks, there should be no problem in designing virtual networking for VMs and hypervisors. The major challenges are the capabilities of security tools and sound designs of network configurations.

1) **Security Tools**. Conventional networking security tools, such as Intrusion Detection System (IDS), Intrusion Prevention System (IPS), are typically running on a physical network and are able to check all of the traffic coming in and out of the area being monitored. They are not capable of examining traffic flowing through internal virtual switches within hypervisors.

2) **Configuration Tools**. Configurations of physical networks can be easily designed by using well-established tools and techniques. But configurations of virtual networks are not easily accessible to network and security professionals making it difficult to validate network designs. In fact, Kim [7] pointed out that a major security risk associated with virtualization is incorrect configuration. It is essential to have a set of best practices in place to ensure that virtual switches and virtual networks are appropriately configured.

# Countermeasures

This section provides practical solutions for hardening virtualized environments.

## A. Hardening Virtual Machines

From a security point of view, virtual machines should be protected just like physical computers. Antivirus applications should be installed and patches should be applied as often as required for physical machines, even if some of those virtual machines remain suspended most of the time, i.e., VMs serving as templates. If applicable, it is better to patch templates and then deploy VMs from templates than patching each VM individually. There should also be a clear organizational policy set up to dictate when to create a new virtual machine so that VM-sprawl problems can be mitigated. Additionally, it is a good idea to schedule tasks from hypervisors for each virtual machine to conduct a full system security scan or a full system backup. It is also advisable to schedule such tasks in a staged manner or during times when physical resources are not heavily utilized.

## B. Protecting Hypervisors

A hypervisor is the foundation of a datacenter and a virtual machine. It must be properly protected from attacks; otherwise, virtual machines will not be able to detect any illegal behavior happening underground. One of the current trends in securing hypervisors is to validate them under the framework of trusted computing. It is advisable to check the integrity of a hypervisor before it can be trusted and deployed [3]. The main challenge here is that trusted computing works only if the underlying hardware, such as CPUs and chipsets, supports Trusted Platform Modules (TPM).

Thus far, no significant security threats directly targeting hypervisors have been discovered. This does not mean that current hypervisor implementations are protected. Attacks on hypervisors tend to be much more difficult than attacking the virtual machines. To counter theoretical attacks such as Subvirt and Blue-Pill threats described earlier more secured virtualization implementations at the chipset level are needed from both Intel and AMD. Another possible solution is to make the hypervisor very thin so that it is easier to validate and can be pre-installed onto the hardware components with read-only capabilities (BIOS-like approach).

## C. Securely Managing Virtual Infrastructure

An effective security policy is a must for any datacenter, whether virtualized or physical. Several levels of permissions can be set in a datacenter. Such levels include datacenter, virtual center, host and virtual–machine, and are represented in Figure 5. The following suggestions are particularly important to enhance the security of virtual infrastructures:

1) **Separate Permissions.** To alleviate the problem of the single point of control, the super administrator, it is best to separate permission policies as the following example indicates [10]:

- Administrator 1 (A1) may reinstall hypervisor OS;
- Administrator 2 (A2) may specify networking for hypervisors and VMs;
- Administrator 3 (A3) may deploy VMs without permissions to modify VMs' local user access and group policies;
- Administrator 4 (A4) manages the distribution of physical resources for a datacenter; and
- A1, A2, A3 and A4 work together as a group to specify administrative tasks for a datacenter.

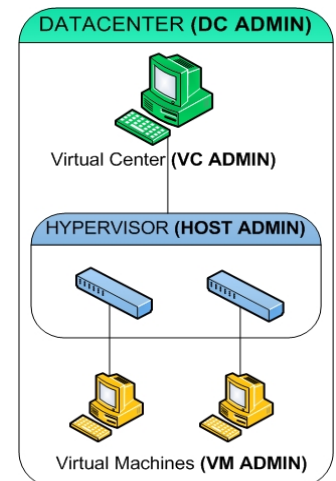

**Figure 5. Levels of permissions and privileges in a virtualized datacenter**

2) **Physical Access**. Physical access to a datacenter is essential: bad practices such as easily accessed codes, access cards and even open doors should be prevented at all times.

3) **Protecting Licensing Servers**. In an enterprise virtual datacenter environment, it is necessary to protect the availability of the licensing server. One solution may be implementing a redundant licensing server.

4) **Securing VM Management Console**. If an attacker gains control of a virtual server management console, either locally or remotely, all hypervisors and VMs that they manage are easily compromised.

## D. Protecting Virtual and Physical Networks

There are crucial steps to be taken when designing a network for a virtualized environment. The following list represents the minimum requirements for configuring networks in virtualized datacenters:

1. Hypervisor management traffic or service console traffic must be restricted and separated from other networks such as virtual machine network, storage (if applicable), etc.
2. Network storage traffic must be separated from virtual machine networks. In other words, network storage access needs to be configured at the hypervisor level, where only the hypervisor has the ability to access datastores and grant access to datastores for VMs. Additionally, SAN administrators need to coordinate access control for hypervisors within datacenters (LUN masking, zoning and others). Most successful secure implementations of Fiber Channel SAN include both hard and soft zoning.
3. Setting security policies on virtual switches is important. Third-party virtual network switches can be used to expand network security.
4. VLAN tagging may be a good option in a limited network resources environment.

Figure 6 depicts a network configuration example for hypervisors.

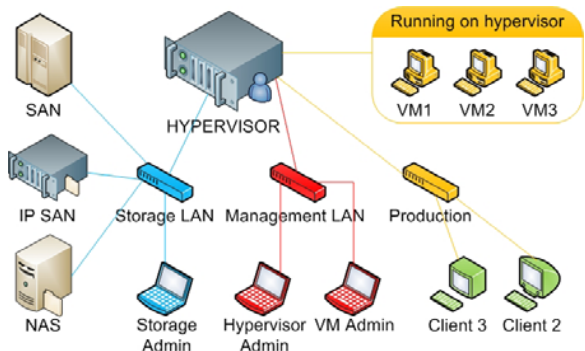

**Figure 6. Separation of network Traffic (without VLANs)**

## E. Recent Virtualization-Aware Security Technologies

Virtualization vendors have recently begun to offer a set of Application Programming Interfaces (APIs) to end users and third-party IT companies. The availability of the hypervisor-level APIs creates an opportunity to build a virtualization-aware security framework because it has the potential to monitor resource usage at both virtual-machine level and physical-machine level. It also makes it possible to examine network traffic passing through the virtual machines and virtual-network interfaces. An overview of such a security platform is shown in Figure 7, where a dedicated security VM is shown being created in a virtual datacenter.

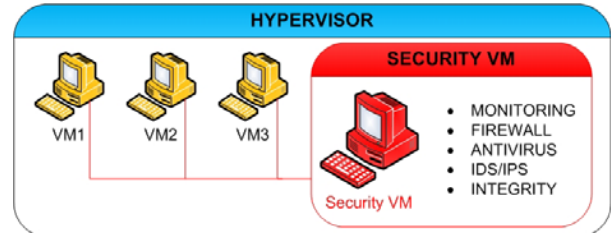

**Figure 7. A model for virtualization security with hypervisor level APIs**

This security VM will be loaded with many security technologies/products such as firewall, Intrusion Detection System (IDS), Intrusion Prevention System (IPS) and anti-virus agents, etc.

The security VM will communicate with the APIs to collect security-related information about each single virtual OS. As a result, it is possible to run a single antivirus agent on the security VM to fully scan multiple virtual VMs. Since hypervisor-level APIs can provide network visibility at the virtual-switch level, traditional firewall and IDS/IPS products can be installed inside the Security VM to protect the virtual datacenter just like they are protecting physical servers. It is the authors' belief that those hypervisor-level APIs can allow the IT industry to quickly arm their virtual datacenters with conventional security tools.

## Conclusions and Recommendations

The current state of virtualization technologies is not more vulnerable than physical servers, but the damage to a virtualized datacenter could be much quicker and more severe than providing services in separate physical environments. However, virtualization technologies present both opportunities and risks. When virtualizing datacenters, all personnel should be involved: server administration teams, networking teams, security, development and management. There should be no difference between protecting VMs and physical computers.

The following may be used as the best practices to enhance virtualization security:

1. Create an effective security policy for the virtualized environment.
2. Define trusted zones and separate servers either at the hardware level or at a VM level.
3. Eliminate single point of control - use separation policies for datacenter administrators.
4. Employ a separation policy for hypervisor administrators and VM administrators.
5. Enforce extremely strict control for virtual center permissions and privileges.
6. In hypervisor cluster settings, provide high availability for license servers and virtual centers – have a primary and a backup copy of Virtual Center and License servers.
7. If no I/O virtualization is deployed, separate physical networks for management and administration of hypervisors, storage, and virtual machines.
8. Disable/remove all unnecessary or superfluous functions and virtual hardware.
9. Prevent virtual machines from utilizing physical resources - do not create and use extra virtual devices such as CPUs, media drives, RAM; never over-allocate any physical resources as this leads to resource-contention problems.
10. Deploy virtual security appliances such as virtual IPS/IDS systems, firewalls, antivirus agents and others in virtualized datacenters.

## Acknowledgments

The authors are grateful to the IJERI reviewers and editors for their valuable feedback and support in the development of this document.

## Biographies

**TIMUR MIRZOEV** in 2003 received the M.S. degree in Electronics and Computer Technology and in 2007 the Ph.D. in Technology Management (Digital Communication) from the Indiana State University. Currently, he is an Assistant Professor of Information Technology at Georgia Southern University. His teaching and research areas include server and network storage virtualization, disaster recovery, storage networks and topologies. Dr. Mirzoev has the following certifications VMware Certified Instructor (VCI), VMware Certified Professional 4 (VCP4), EMC Proven Professional, LefthandNetworks (HP) SAN/iQ, A+. He may be reached at tmirzoev@georgiasouthern.edu

**BAIJIAN YANG** received the Ph.D. in Computer Science from Michigan State University in 2002. Currently, he is an Assistant Professor in the Department of Technology at Ball State University. His teaching and research areas include Information Security, Distributed Computing, Computer Networks, and Server Administration. He is also a Certified Information System Security Professionals (CISSP). Dr. Yang can be reached at byang@bsu.edu